# Photoemission response of 2D electron states

V.N. Strocov

*Swiss Light Source, Paul Scherrer Institute, CH-5232 Villigen-PSI, Switzerland*

*(vladimir.strocov@psi.ch)*

A lucid Fourier analysis based description of the photoemission process is presented that directly relates photon energy (*hv*) dependent ARPES response of two-dimensional (2D) electron states to their wavefunctions. The states formed by quantum confinement of bulk Bloch waves (including Shockley-Tamm type surface and interface states, and quantum-well states) show periodic peaks of ARPES intensity as a function of *hv*. Amplitudes of these peaks reflect Fourier series of the oscillating Bloch-wave component of the wavefunction, and their broadening spatial confinement of its envelope function. In contrast, the 2D formed by local orbitals (dangling bonds and defects at the surface or interface) show aperiodic *hv*-dependence, where the rate of decay reflects localization of these states in the out-of-plane direction. This formalism sets up a straightforward methodology to access fundamental properties of different 2D states, as illustrated by analysis of previous photoemission experimental data including the paradigm Al(100) surface state, quantum-well states in multilayer graphene and at the buried GaAlN/GaN interface, and molecular orbitals.

# Introduction

Two-dimensional (2D) electron states formed by quantum confinement (QC) play an important role in fundamental physics of solid-states systems and their device applications. One of numerous examples is the GaAlAs/GaAs interface where confinement of mobile electrons in a quantum-well (QW) formed by the interfacial band bending allowed realization of high electron mobility transistors (HEMTs) presently used in virtually all millimetre and microwave communication devices including cell phones (Mimura 2002). Such QW-states are formed by confinement of oscillating bulk Bloch waves whose out-of-plane periodicity is imposed by the periodic bulk potential $V(\mathbf{r})$ (for entries see (Chiang 2000 and Yoshimatsu *et al* 2011). Surface states of the Shockley-Tamm type is another example of electron states which can be viewed as resulting from QC, in this case between vacuum and a band gap in crystal bulk (Heine 1963, Inglesfield 1982). Their wavefunction is an evanescent Bloch wave whose periodic $B(z)$-part is inherited from the bulk Bloch waves at the band gap edges (for theory of complex band structure see, for example, Heine 1963, Dederichs 1972 and Reuter 2017). Here, all such states formed by confinement of periodic bulk Bloch waves are treated on the same footing using a generalized terminology of *QC-type* states. The wavefunction of these states (sketched later on in Fig. 1) is represented by an oscillating Bloch wave, inherent in the periodic bulk $V(\mathbf{r})$, which is modulated by an envelope function describing the confinement. Another type of 2D states is local-orbital type (*LO-type*) states such as surface and interface states induced by dangling bonds or defects (Inglesfield 1982). Although the former can be delocalized in the in-plane direction and the latter not, common to all such states is that they are localized in the out-of-plane direction and not connected with confinement of bulk states, with their wavefunction (also sketched in Fig. 1) essentially reduced to the aperiodic envelope function. As we will see here, the presence or absence of the oscillating wavefunction component causes totally different photoemission response of these two types of 2D states.

Angle-resolved photoelectron spectroscopy (ARPES) is the unique experimental method that allows explore of electron states with resolution in their binding energy $E_b$ and momentum $\mathbf{k}$ (see, for example, the recent review by Suga & Tusche 2015). Surface states have in the past been one of the most popular applications of ARPES in the VUV photon energy range ($h\nu$ < 100 eV) where the electron mean free path $\lambda$, defining the probing depth of this technique, is of the order of 0.5 nm (Powell *et al* 1999). $\mathbf{k}$-resolved studies of electron states at buried interfaces have however become accessible only with advent ARPES in the soft-X-ray range ($h\nu$ typically from 300 to 1500 eV) where increase of photoelectron kinetic energy $E_k$ boosts $\lambda$ by a factor of 3-5 (Strocov *et al* 2014, Cancellieri *et al* 2016).

Owing to 2D character of all surface and interface states, their ARPES response shows spectral peaks whose $E_b$ is independent of out-of-plane photoelectron momentum $K_z$ varied in the ARPES experiment through $h\nu$. Characteristic of the QC-type states is however that the corresponding ARPES intensity periodically oscillates as a function of $K_z$. First noted in the pioneering work of Louie *et al* (1980) on the Cu(111) surface state, this phenomenon was later studied experimentally and theoretically for a number of QC-systems including metal thin films (Ortega *et al* 1993, Mugarza *et al* 2001, Chiang *et al* 2000), multilayer graphene (Ohta *et al* 2007), buried delta-layers [Miwa *et al* 2013] and various surface states (Kevan *et al* 1985, Hofmann *et al* 2004, Krasovskii *et al* 2008, Borghetti *et al* 2012). The LO-type states, in contrast, show an aperiodic ARPES response decaying with $K_z$.

Following the previous works [Louie *et al* 1980, Kevan *et al* 1985], a lucid formalism is developed here that directly relates the ARPES intensity oscillations to Fourier series and spatial confinement of the 2D states, and can therefore be used for their experimental determination. Of crucial importance for this purpose is the use of high excitation energies whereby the ARPES final states become pure plane waves without their hybridization (Strocov *et al* 2012). This formalism is illustrated in application to previous photoemission experimental data on the Al(100) surface (Hofmann *et al* 2004).

## Fourier-transform formalism of ARPES

Within the one-step theory of photoemission – see, for example, Feibelmann & Eastman (1974) – the ARPES intensity is found as $I_{PE} \propto |\langle f|\mathbf{A}\cdot\mathbf{p}|i\rangle|^2$, where $\langle f|$ is the final state described as the time-reversed LEED state and $|i\rangle$ the initial state, which are coupled through the vector potential $\mathbf{A}$ of incident electromagnetic field and momentum operator $\mathbf{p}$. Neglecting the explicit matrix element effects embedding the experimental geometry (see, for example, Moser 2017), this expression can be simplified to the scalar product of $\langle f|$ and $|i\rangle$, $I_{PE} \propto |\langle f|i\rangle|^2$. We here assume that the final-state momentum $\mathbf{K}$ is corrected back to the initial-state one $\mathbf{k}$ for the photon momentum $p = h\nu/c$ which is significant in the soft-X-ray range.

For sufficiently high excitation energy $h\nu$, when $E_k$ much exceeds the $V(\mathbf{r})$ corrugation, the free-electron (FE) approximation usually holds for the final states which can in this case be represented by a plane wave $e^{i\mathbf{Kr}}$ periodic in the in-plane direction $\mathbf{r}_{xy}$ and damped in the out-of-plane one $z$ (non-FE effects in the final states, which can be particularly strong at low excitation energies $h\nu$ – see Strocov *et al* (2006) and references therein – will be discussed later). Introducing out-of-plane damping of final state due to incoherent scattering processes, $\langle f|$ can be represented as $\langle f| \propto e^{i\mathbf{K}_{xy}\mathbf{r}_{xy}} e^{iK_z z}$, where $K_z$ is complex, with its real part $K_z^r$ describing the oscillating $z$-dependent part of the $\langle f|$-wavefunction and imaginary part $K_z^i$ its damping into the sample depth due to finite photoelectron mean free path $\lambda = 1/2K_z^i$ (the factor 2 comes from squaring of the wavefunction amplitude for electron density), $\langle f| \propto e^{i\mathbf{K}_{xy}\mathbf{r}_{xy}} e^{iK_z^r z} e^{-K_z^i z}$. In turn, $|i\rangle$ also obeys the in-plane periodicity, and can be expanded over 2D reciprocal vectors $\mathbf{g}$ as $|i\rangle = \sum_{\mathbf{g}} A_{\mathbf{k}_{xy}+\mathbf{g}}(z) e^{i(\mathbf{k}_{xy}+\mathbf{g})\mathbf{r}_{xy}}$, where $\mathbf{k}_{xy}$ is the in-plane electron momentum in the reduced BZ and coefficients $A_{\mathbf{k}_{xy}+\mathbf{g}}(z)$ represent the $|i\rangle$-wavefunction behavior in the $z$-direction, being periodic for 3D states and confined for 2D states. Orthogonality of the in-plane propagating plane waves in the above expansions for $\langle f|$ and $|i\rangle$ implies that the only term in $|i\rangle$ giving non-zero contribution to $\langle f|i\rangle$ satisfies the condition $\mathbf{k}_{xy} + \mathbf{g} = \mathbf{K}_{xy}$ i.e. the above sum over $\mathbf{g}$ reduces to $|i\rangle = A_{\mathbf{K}_{xy}}(z) e^{i\mathbf{K}_{xy}\mathbf{r}_{xy}}$.

With the above expansions of $\langle f|$ and $|i\rangle$, $I_{PE}$ as a function of $\mathbf{K}_{xy}$ becomes $I_{PE}(\mathbf{K}_{xy}) \propto |\langle e^{i\mathbf{K}_{xy}\mathbf{r}_{xy}} e^{iK_z^r z} e^{-K_z^i z} | A_{\mathbf{K}_{xy}}(z) e^{i\mathbf{K}_{xy}\mathbf{r}_{xy}} \rangle|^2 = |\langle e^{iK_z^r z} e^{-K_z^i z} | A_{\mathbf{K}_{xy}}(z) \rangle|^2$. Being proportional to $|A_{\mathbf{K}_{xy}}|^2$, the

$\mathbf{K}_{xy}$-dependent $I_{PE}$ reflects essentially the 2D Fourier series of the $|i\rangle$-states periodic in the out-of-plane spatial coordinates.

Transferring the damping $e^{-K_z^i z}$ into the $|i\rangle$-part, we obtain $I_{PE}^{\mathbf{K}_{xy}} \propto \left| \left\langle e^{iK_z^r z} \left| e^{-K_z^i z} A_{\mathbf{K}_{xy}}(z) \right\rangle \right|^2$. In the explicit form, this expression reads as $I_{PE}^{\mathbf{K}_{xy}}(K_z) \propto \left| \int_0^\infty e^{iK_z^r z} \left\{ e^{-K_z^i z} A_{\mathbf{K}_{xy}}(z) \right\} dz \right|^2$, where the lower integration limit implies that the $|i\rangle$-wavefunction terminates at the surface $z=0$. Physically, this formula shows that the $K_z$-dependent $I_{PE}$ for given photoelectron $\mathbf{K}_{xy}$ is simply the squared FT

$$I_{PE}^{\mathbf{K}_{xy}}(K_z) \propto \left| F_{K_z} \left\{ e^{-K_z^i z} A_{\mathbf{K}_{xy}}(z) \right\} \right|^2 \tag{1}$$

of the (final-state damping weighted) $z$-dependent coefficient $A(z)$ for this $\mathbf{K}_{xy}$ in 2D expansion of the $|i\rangle$-wavefunction. This property of the ARPES response can in principle be used to reconstruct the wavefunctions of 2D states using iterative algorithms similar to those used in molecular wavefunction reconstruction (Puschnig *et al* 2009, Weiss *et al* 2015, Bradshaw & Woodruff 2015, Kliuiev *et al* 2016). If $\lambda$ is much larger than the $|i\rangle$-wavefunction localization region, a likely situation for sufficiently high *hv*, the above FT simplifies to $I_{PE}^{\mathbf{K}_{xy}}(K_z) \propto \left| F_{K_z} \left\{ A_{\mathbf{K}_{xy}}(z) \right\} \right|^2$. We note that whereas the $\mathbf{K}_{xy}$-dependence of $I_{PE}$ is represented by the Fourier series, whose discrete character reflects the in-plane periodicity of the system, its $K_z$-dependence is represented by the integral Fourier transform, whose continuous character reflects the out-of-plane aperiodicity.

We will now focus on the 2D states that are confined in the out-of-plane direction. For the QC-type states sketched in Fig. 1 (*a*), the coefficients $A_{\mathbf{K}_{xy}}(z)$ representing the $|i\rangle$-wavefunction for given $\mathbf{K}_{xy}$ can be written as a slowly varying envelope function $E(z)$ that modulates a Bloch wave $B_{\mathbf{K}_{xy}}(z)$ which is derived from the periodic bulk $V(\mathbf{r})$, periodic in the out-of-plane direction and characterized by certain momentum $k_z$ (Ortega *et al* 1996, Chiang 2000, Mugarza *et al* 2001)

$$A_{\mathbf{K}_{xy}}(z) = E(z) \cdot B_{\mathbf{K}_{xy}}(z) \tag{2}$$

We note that the $E(z)$-functions can described analytically by Airy functions, if $V(z)$ confining the QC-type states is approximated by a triangular shape, and by Bessel functions if a more appropriate exponential approximation $V(z) \propto -V_0 e^{-2z/a}$ is used, where $V_0$ is the characteristic depth and $a$ width of $V(z)$ (Jovic *et al* 2017, Moser *et al* 2018). The $B(z)$-functions, strictly speaking, are standing waves including two Bloch waves with $\pm k_z$ propagating in opposite out-of-plane directions, however, for brevity of the formalism we keep here only one of them. The same expression (2) applies to the LO-type states also sketched in Fig. 1 (*b*), and in that case $A_{\mathbf{K}_{xy}}(z)$ reduces to $E(z)$, with $B_{\mathbf{K}_{xy}}(z) = B_{\mathbf{K}_{xy}}$ remaining merely a constant depending on $\mathbf{K}_{xy}$.

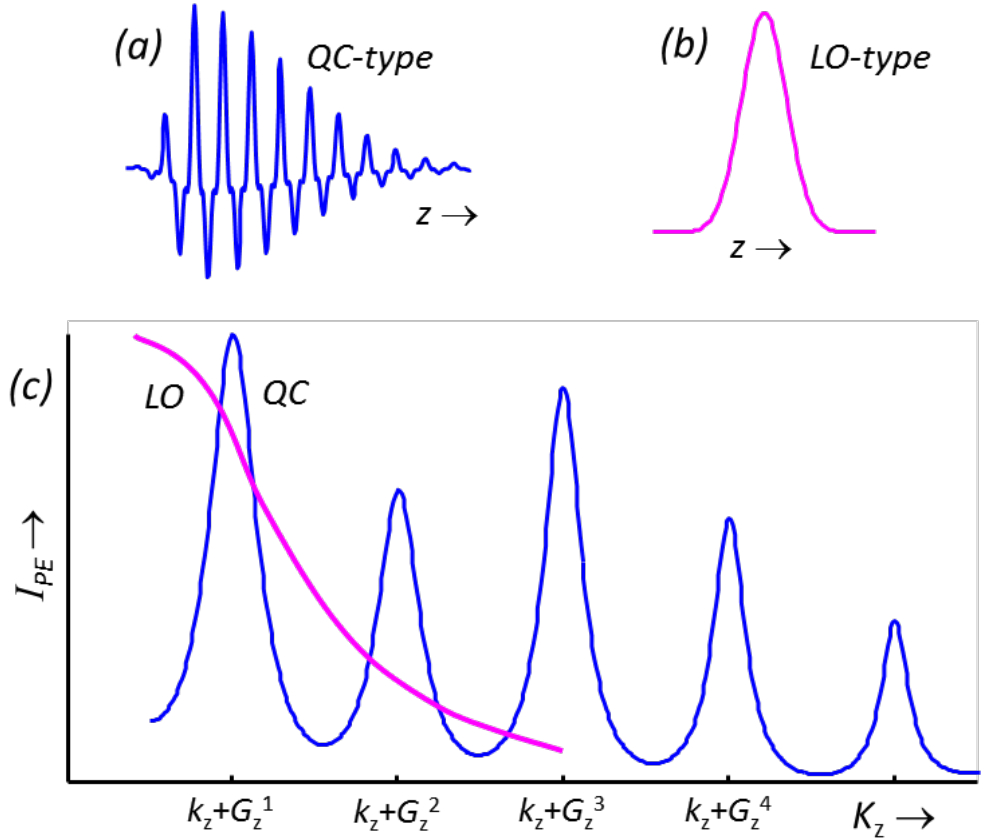

**Fig. 1.** Schematic wavefunctions of the QC-type (*a*) and LO-type (*b*), and the corresponding pattern of their ARPES response $I_{PE}(K_z)$ (*c*). Due to the higher harmonics in the Bloch-wave part of the QC-type states, their ARPES response is characterized by $G_z$-periodic peaks whose amplitudes reflect Fourier series of the periodic Bloch-wave part of these states, and widths their spatial extension combined with the photoelectron λ. In contrast, the LO-type states are characterized by an aperiodic decay of $I_{PE}(K_z)$. In any case the $I_{PE}(K_z)$-dependences vary with $\mathbf{K}_{xy}$.

In general case, the $B_{\mathbf{K}_{xy}}(z)$ adapted to the periodic *V*(**r**) can be expanded in a Fourier series over out-of-plane reciprocal vectors $G_z$ of the host 3D lattice as $B_{\mathbf{K}_{xy}}(z) = \sum_{G_z} C_{\mathbf{K}_{xy}}^{G_z} e^{i(k_z+G_z)z}$. Assembling all terms in the above expression (1) for the photocurrent $I_{PE}^{\mathbf{K}_{xy}}(K_z)$, we arrive at $I_{PE}^{\mathbf{K}_{xy}}(K_z) \propto$

$$\left| F_{K_z}\left\{ e^{-K_z^i z} \cdot E(z) \cdot \sum_{G_z} C_{\mathbf{K}_{xy}}^{G_z} e^{i(k_z+G_z)z} \right\} \right|^2 = \left| \sum_{G_z} C_{\mathbf{K}_{xy}}^{G_z} F_{K_z}\left\{ e^{-K_z^i z} \cdot E(z) \cdot e^{i(k_z+G_z)z} \right\} \right|^2 =$$

$$\left| \sum_{G_z} C_{\mathbf{K}_{xy}}^{G_z} F_{K_z}\left\{ e^{-K_z^i z} \cdot E(z) \right\} * F_{K_z}\left\{ e^{i(k_z+G_z)z} \right\} \right|^2,$$ where the FT is broken down in convolution of two transforms. As $F_{K_z}\left\{ e^{i(k_z+G_z)z} \right\}$ equals simply to the δ-function δ($K_z$-($k_z$+$G_z$)), this yields

$I_{PE}^{\mathbf{K}_{xy}}(K_z) \propto \left| \sum_{G_z} C_{\mathbf{K}_{xy}}^{G_z} F_{K_z} \left\{ e^{-K_z^i z} \cdot E(z) \right\} * \delta(K_z - (k_z + G_z)) \right|^2$. Using the δ-function convolution rule $f(x)*\delta(x-a) = f(x-a)$, we obtain the following expression for the $K_z$-dependent ARPES response $I_{PE}(K_z)$ of 2D states:

$$I_{PE}^{\mathbf{K}_{xy}}(K_z) \propto \left| \sum_{G_z} C_{\mathbf{K}_{xy}}^{G_z} F_{K_z - (k_z + G_z)} \left\{ e^{-K_z^i z} \cdot E(z) \right\} \right|^2 \quad (3)$$

Inside the squared absolute value, this expression sums up over a series of functions $F_{K_z - (k_z + G_z)} \left\{ e^{-K_z^i z} \cdot E(z) \right\}$. As the ⟨f|- and 2D |i⟩-wavefunctions both decay with z, these functions are in general peaks centered at $k_z + G_z$ which decay with detuning of $K_z$ from the peak center.

Essential for the validity of this straightforward connection between the ARPES response and Fourier composition of the |i⟩-states is that the final states have the FE-character, i.e. described by a pure plane wave $e^{i\mathbf{Kr}}$. Otherwise, these states are a mixture of plane waves $\sum_{\mathbf{K}} A_{\mathbf{K}} e^{\mathbf{Kr}}$ extending over **K** with different $\mathbf{K}_{xy}$ and $K_z$ projections. In this case $I_{PE}^{\mathbf{K}_{xy}}(K_z)$ would mix up the corresponding Fourier components of the |i⟩-state, rendering interpretation of the experimental data ambiguous. In general, high hv much exceeding the atomic corrugation of V(**r**) are required to ensure purely FE-character of the final states (Strocov et al 2012).

The above lucid description of the ARPES process obviously neglects the experimental geometry and photon polarization effects as well as non-FE effects in the final states important for quantitative description of the ARPES intensity. They are correctly taken into account by the one-step photoemission theory within its Bloch-wave or scattering theory formulations, for entries relevant for the 2D states see Krasovskii & Schattke 2004, Krasovskii et al 2008, and Braun & Donath 2004, respectively. The heavy computational machinery of these first-principles methods obscures however the Fourier composition roots of the ARPES intensity behavior.

## Characteristic ARPES response patterns

### QC-states

The QC-type states, sketched in Fig. 1 *(a)*, typically have a long-range character expressed by relatively large extension of their $E(z)$. If the final-state extension is also relatively large as expressed by small $K_z^i$, which is typical of high $E_k$, the peaks of the $F_{K_z - (k_z + G_z)} \left\{ e^{-K_z^i z} \cdot E(z) \right\}$ functions in the above expression (3) for ARPES intensity do not overlap in $K_z$. In this case the squared absolute value can be exchanged with the sum as

$$I_{PE}^{\mathbf{K}_{xy}}(K_z) \propto \sum_{G_z} \left| C_{\mathbf{K}_{xy}}^{G_z} \right|^2 \left| F_{K_z - (k_z + G_z)} \left\{ e^{-K_z^i z} \cdot E(z) \right\} \right|^2. \quad (4)$$

This formula contains the key mathematical message of this work: $K_z$-dependent ARPES response of the QC-states appears as a sequence of $G_z$-periodic peaks weighted by the Fourier series amplitudes of the underlying Bloch wave. This characteristic pattern is sketched in Fig. 1 (*c*). We should highlight its three important aspects:

(1) The peaks are centered at the $K_z$-values equal to the $k_z+G_z$ vectors of the Fourier series of the periodic $B(z)$-part of the QC-wavefunction (Louie *et al* 1980, Ortega *et al* 1993, Mugarza *et al* 2000). This fact is a manifestation of the periodic Bloch wave origin of the QC-state. Physically, whenever $K_z$ of the ⟨*f*|-wavefunction hits some $k_z+G_z$ harmonic of the |*i*⟩-one, the ARPES intensity blows up. In the reduced Brillouin zone picture, this condition is equivalent to the out-of-plane momentum conservation $K_z = k_z$ (vertical transitions) between the ⟨*f*|- and |*i*⟩-states.

As a minor refinement to this picture, we recall that the standing-wave-like QC-type states in principle include two Bloch waves with opposite signs of $k_z$, splitting each $I_{PE}(K_z)$ peak into two ones separated by $2|k_z|$, but this separation is usually much smaller than $G_z$. Furthermore, we note that the theoretical approaches which reduce the QC-type states to merely $E(z)$-functions neglecting the periodic $B(z)$-parts (Jovic *et al* 2017, Moser *at al* 2018) are unable to reproduce the characteristic $K_z$-periodicity of $I_{PE}$. Our formalism reduces to these approaches by retaining in the sum (4) only one non-zero $C_{\mathbf{K}_{xy}}^{G_z}$ coefficient at $G_z = 0$.

(2) The peak amplitudes are proportional to $\left|C_{\mathbf{K}_{xy}}^{G_z}\right|^2$ defined by the Fourier-series coefficients of the $B(z)$-functions. Therefore, the $K_z$-dependent ARPES intensity informs the Fourier series of the periodic Bloch-wave content of the QC-wavefunction.

Furthemore, we note that the peak values at $K_z = k_z + G_z$ are all proportional to $\left|F_0\left\{e^{-K_z^i z} \cdot E(z)\right\}\right|^2$ or, in the explicit form, to $\left|\int_0^\infty e^{-K_z^i z} E(z) dz\right|^2$. This means that the ARPES signal scales up with spatial extension of both ⟨*f*|- and |*i*⟩-states into the crystal interior. The $I_{PE}(K_z)$-dependences vary with $\mathbf{K}_{xy}$ through the $\left|C_{\mathbf{K}_{xy}}^{G_z}\right|^2$ coefficients.

(3) The peaks have the shape defined by FT of the $E(z)$ envelope function of the |*i*⟩-states weighted with the final-state decay $e^{-K_z^i z}$. With the former being the same for all peaks, the peak shape will however depend on $K_z$ through the $\lambda(E_k)$ dependence. Therefore, $K_z$-broadening of the ARPES peaks informs the $E(z)$-function of the 2D valence electron states. The correct shape of the $I_{PE}(K_z)$ peaks is reproduced even theoretical approaches neglecting the periodic $B(z)$-content of the QC-wavefunctions (Jovic *et al* 2017, Moser *at al* 2018).

Unfolding the last point, we note that in many cases, in particular for the surface states, $E(z)$ can be approximated by exponent $e^{-Dz}$, where *D* is the decay constant connected with the attenuation length *d* as $D=1/2d$ (the factor 2 comes from squaring of the wavefunction amplitude for electron density). Combining it with the final-state damping $e^{-K_z^i z}$, we obtain $\left| F_{K_z-(k_z+G_z)} \left\{ e^{-K_z^i z} \cdot E(z) \right\} \right|^2 =$ $\left| F_{K_z-(k_z+G_z)} \left\{ e^{-(K_z^i+D)z} \right\} \right|^2$. The latter is a Lorentzian $\left( 1 + \left( \frac{K_z-(k_z+G_z)}{K_z^i + D} \right)^2 \right)^{-1}$ having full width at half maximum (FWHM) equal to $\Delta = 2(K_z^i + D)$ ($E_k$-dependence of $K_z^i = 1/2\lambda$ over the peak width can be neglected). In terms of attenuation length, the obtained FWHM is

$$\Delta = \lambda^{-1} + d^{-1} \tag{5}$$

Therefore, with a good estimate of photoelectron $\lambda$ for given $E_k$, Lorentzian FWHM of an experimental $I_{PE}(K_z)$-peak directly informs spatial extension of the QC-state. Under the condition $\lambda \gg d$, which might be achieved at sufficiently high $h\nu$, the FWHM will be simply equal to $d^{-1}$. We note that the QC-wavefunctions described by Bessel functions in exponential $V(z)$ deliver the ARPES peaks somewhat different in shape and width (Jovic *et al* 2017, Moser *et al* 2018) compared to the above $e^{-Dz}$ wavefunctions.

We note that the above FT-based formalism puts on rigorous grounds and generalizes similar earlier theories to describe ARPES of surface states (Louie *et al* 1980, Kevan *et al* 1985). We note that often the QC-states are described within a simplified phase accumulation model that essentially replaces the sharply oscillating Bloch waves by smooth plane waves (see, for example, the review by Milun *et al* 2002) or simply reducing them to solely the $E(z)$-function confined in one-dimensional $V(z)$ (Mooser *et al* 2018 and references). In a sense, this reminds the idea of pseudopotential where the real sharply oscillating wavefunction is replaced by its smooth pseudo-wavefunction analogue. Similarly, the phase accumulation model is only relevant for the energy levels and $E(z)$-functions. However, neglecting the effects embedded in the periodic $B(z)$-functions, this model is unable to correctly reproduce the characteristic periodic ARPES response of the QC-states.

### *LO-states*

We will now turn to another type of 2D states, the LO-type ones, which have purely surface or interface origin and are unrelated to any bulk Bloch waves. Schematically shown in Fig. 1 (*b*), such out-of-plane localized states can be formed, for example, by dangling bonds on semiconductor surfaces (Inglesfield 1982) or by surface and interface defect states. $I_{PE}(K_z)$-dependence for such states follows a pattern totally different from the above QC-states. In this case the sum over $G_z$ in the above expression (3) for ARPES intensity retains the sole term

$$I_{PE}^{\mathbf{K}_{xy}}(K_z) \propto \left| B_{\mathbf{K}_{xy}} \right|^2 \left| F_{K_z} \left\{ e^{-K_z^i z} \cdot E(z) \right\} \right|^2, \tag{6}$$

where $C_{\mathbf{K}_{xy}}^{G_z=0}$ is replaced back by the $\mathbf{K}_{xy}$-dependent constant $B_{\mathbf{K}_{xy}}$ from the expression (2). Essentially, ARPES response of the LO-type states is reduced to merely the first $G_z = 0$ peak of the $I_{PE}(K_z)$ dependence (3).

In contrast to the QC-type states, such $I_{PE}(K_z)$-dependence is aperiodic and, as $E(z)$ is typically smooth and dominated by low spatial frequencies, rapidly decays with $K_z$ as sketched in Fig. 1 (*c*). This characteristic ARPES response of the LO-type states sharply contrasts to the periodic and slowly decaying ARPES response of the QC-type states. We note that in the LO-case the $I_{PE}(K_z)$ dependence varies with **K**$_{xy}$ through the coefficients $B_{\mathbf{K}_{xy}}$.

If wavefunctions of the LO-states are symmetric in the out-of-plane direction like the *s*- or $p_{xy}$-orbitals, in many cases their $E(z)$ can be approximated by a Gaussian $e^{-2\ln 2 \left(\frac{z}{d}\right)^2}$ where $d$ is FWHM of corresponding electron density distribution (mind the factor 2 coming from the wavefunction squaring). Combining it with the final-state damping $e^{-K_z^i z}$, for any **K**$_{xy}$ we obtain

$$I_{PE}^{\mathbf{K}_{xy}}(K_z) \propto \left| F_{K_z} \left\{ e^{-K_z^i z} \cdot e^{-2\ln 2 \left(\frac{z}{d}\right)^2} \right\} \right|^2 = \left( 1 + \left(\frac{K_z}{K_z^i}\right)^2 \right)^{-1} * e^{-\frac{d^2}{8\ln 2}K_z^2}$$

which is a convolution of Lorentzian (distorted by the $K_z$-dependence of $K_z^i$) and Gaussian profiles centered at $K_z = 0$. Physically, the smaller $\lambda$ and $d$ in these profiles, the faster the decay of this ARPES intensity with *hv*. With a good estimate of photoelectron $\lambda(E)$, fitting of the experimental $K_z$-dependence with this profile will yield an estimate of the wavefunction spatial extension *d*. If we can neglect the $K_z$-dependence of $K_z^i$ through the Lorentzian profile, $I_{PE}(K_z)$ reduces to a Voigt profile with its FWHM approximated as

$$0.53 K_z^i + \sqrt{0.22 \left(K_z^i\right)^2 + \left(\frac{4\sqrt{2\ln 2}}{d}\right)^2}.$$

Accuracy of this approximation improves with decrease of *d* towards $d \ll \lambda$, when the Lorentzian component with less certain $K_z^i$ becomes negligible. We note that whereas these formulas, based on Gaussian approximation of the out-of-plane symmetric LO-wavefunctions, predict the increase of $I_{PE}(K_z)$ towards $K_z = 0$, for the antisymmetric wavefunctions like the $p_z$-orbitals the corresponding $I_{PE}(K_z)$ goes there to zero (Weiss *et al* 2015). In any case, however, $I_{PE}(K_z)$ is aperiodic and decays with increase of $K_z$.

## Analysis of experimental ARPES data

### *QC-states*

Belonging to the QC-type states are the Shockley-Tamm surface states which are essentially bulk Bloch waves, confined at the surface by band gaps in the bulk band structure. Owing to the higher Fourier components of the Bloch waves and typically long-range character of these surface states, the oscillating ARPES signal from them can persist with increase of *hv* up to 1 keV and even higher. This can be illustrated by results of Hofmann *et al* (2002) on the Shockley-Tamm surface state on Al(100) which are reproduced in Fig. 2. The ARPES peaks coming from the bulk *sp*-band disperse as a function of $K_z$ varied through *hv*. The dispersion maxima at three consecutive *hv*-values are achieved where $K_z$ reaches the X-point. The surface state peak (marked *SS*) in the *sp*-band gap, in contrast, does not disperse as a function of $K_z$ that is characteristic of its 2D nature. However, its intensity blows up whenever $K_z$ reaches the X-point, following the periodic $I_{PE}(K_z)$ pattern in Fig. 1 (c). This indicates that $k_z$ of the periodic $B(z)$-part of the surface state wavefunction falls onto the X-point. Even at high hv the ARPES response of the surface state stays comparable with that of the bulk states, which is consistent with its large out-of-plane extension of >10 Å. However, a quantitative evaluation of E(z) from the these data is difficult because energy resolution of this early soft-X-ray ARPES experiment was insufficient to resolve the surface state peak from the sp-bands in the X-point. Another example of strong oscillating ARPES response persisting at high hv is the surface state in BiTeI (Landolt *et al* 2013).

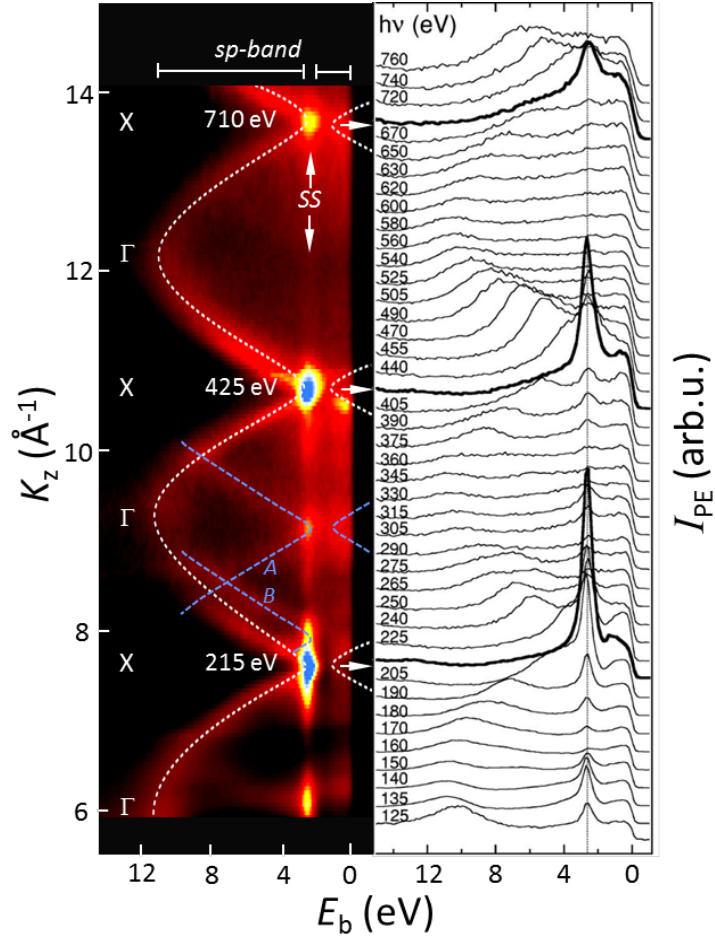

**Fig. 2.** ARPES data on Al(100) in a wide hv-range. The ARPES peaks oscillating in $E_b$ as a function of hv manifest the bulk sp-bands, and those at constant $E_b$ the surface state (SS). In agreement with the characteristic pattern in Fig. 1, its response blows up whenever $K_z$ comes to the X-point. The secondary dispersion branches A and B of the sp-bands manifest multiband composition of the final states (adapted from Hofmann *et al* 2002).

An intriguing peculiarity of the data on Al(100) is that apart from the main dispersion branch of the bulk ARPES peaks one can observe few weaker secondary branches (marked A-B) that correspond to different $k_z$-values. In particular, hv ~ 330 eV brings $k_z$ of the main branch to the Γ-point and that of the branch A to X. This fact evidences that in this case the final state deviates from the FE-type plane wave $e^{i\mathbf{Kr}}$ that would have selected one single $k_z$. In fact, it incorporates two different bands with different $k_z$ and leading Fourier components $e^{i(k_z+G_z)z}$ which produce ARPES peaks of similar intensity. One may think of such multiband final-state as a composition of umklapp bands or, in the Mahan's terminology, secondary photoemission cones (Mahan 1970). This phenomenon beyond the simple FE-approximation, resulting from hybridization of plane waves through $V(\mathbf{r})$, has been studied experimentally and theoretically in low-energy ARPES for bulk bands of various materials including Cu (Strocov *et al* 1997) and transition metal dichalcogenides (Strocov *et al* 1997, Strocov *et al* 2006, Krasovskii *et al* 2007) as well as surface states, in particular for the Al(100) and (111) surfaces (Krasovskii *et al* 2008). It is surprising to see in Fig. 2 that, although Al is one of the most FE-materials, the multiband final-state composition carries on at least to 400 eV. This phenomenon is actually surprisingly common in the soft-X-ray energy range, and will be addressed in detail elsewhere. Here we only note that it does not in general hamper interpretation of soft-X-ray ARPES data, because the intrinsic out-of-plane momentum broadening $\Delta k_z$ (Strocov 2003) in this $h\nu$-range typically reduces below $k_z$-separation of different Bloch waves that allows reliable assignment of the corresponding spectral structures. Non-FE effects in ARPES of molecular systems are discussed below.

Further examples of $h\nu$-dependent ARPES of surface states, starting from the pioneering work of Louie *et al* (1980) on Cu(111), include various surfaces of Al, Cu, Ag and Au (Kevan *et al* 1985, Krasovskii *et al* 2008, Borghetti *et al* 2012), topological materials (Bianchi *et al* 2010), Weil semimetals (Jiang *et al* 2017), etc.

For the QW-states, an impressive example of their periodic ARPES response has been demonstrated for thin Ag films on Ni(111) (Miller *et al* 1994) where constant-initial-state spectral peaks from the states confined in Ag exposed a clear cyclic behaviour of their intensity. Another example is thin Cu(100) films on Co(100) (Mugarza *et al* 2000, Mugarza *et al* 2001). In agreement with the pattern in Fig. 1, the ARPES peaks have been found at $h\nu$-values corresponding to vertical transitions between the initial-state $k_z$ and final-state $K_z$. Furthermore, this study has identified a multiband final-state composition at the Cu(100) surface going up to 100 eV (Strocov *et al* 1997, Mugarza *et al* 2001). Another study of QW-states formed in Cu films on Co(100) and Co(110) (Hansen *et al* 1997) has shown that modulations of their $h\nu$-dependent $I_{PE}(h\nu)$ are related by $k_z$ conservation to dispersion of the Cu bulk *sp*-band that these QW-states are derived from, in agreement with our FT-based theory. Semiconductor surfaces can also bear QW-states formed from bulk CB states confined in the surface band-bending $V(\mathbf{r})$. A recent example is InN where the band bending can be tuned, for example, by absorption of potassium (Colakerol *et al* 2015) or water (Jovic *et al* 2017). These QW-states demonstrate the same periodic ARPES response pattern (Jovic *et al* 2017). We note that at sufficiently low $E_k$ the QW can also confine the $\langle f|$-wavefunctions just as the $|i\rangle$-ones. This effect introduces additional modulations to the ARPES spectra (Paggel *et al* 1999, Chiang 2000).

An illuminating ARPES study of QW-states in multilayer graphene films, tracing their transformation from 2D to three-dimensional (3D) character with increase of the film thickness, has been reported by Ohta *et al* (2007). In this work the formation of QW-states was described through multilayer splitting of the π-states coming from each graphene sheet, an approach complementary to the conventional one based on the confinement of delocalized Bloch waves. The ARPES results from this work are reproduced in Fig. 3. The panels (*a-d*) show the 2D bands near $E_F$ which are formed in a stack of $m$ = 1 - 4 graphene layers due to multilayer splitting of the π- and π*-states from each layer. The corresponding $K_z$-dependent ARPES response of the π-bands at $E_F$ - 1 eV is shown in (*e-h*). For the monolayer (*m*=1) graphene, an aperiodic $I_{PE}(K_z)$ slowly decaying with $h\nu$ is observed. This reflects, in accordance with our theoretical picture, the LO-type character of the π*-state in the single monolayer. As $m$ progressively increases to 4 layers, the π-states split into $m$ pairs of states confined within the stack and, owing to the out-of-plane mirror symmetry of the system, having symmetric $\pm k_z$ values. The corresponding $I_{PE}(K_z)$ evolves to a sequence of peak pairs whose period corresponds to the interlayer distance, exactly the pattern our FT-based formalism predicts for the QC-type states. The peak width progressively decreases with $m$, reflecting the increase of electron delocalization over the graphene multilayer, expressed by the $E(z)$-function. With increase of $m$ the individual multilayer-split 2D states form, remarkably, a constellation of peaks whose pattern in the $(k_x, K_z)$ coordinates gradually approaches a $K_z$-dispersive pattern of ARPES intensity characteristic of the 3D states. This mechanism of the 2D to 3D transformation of the ARPES response, based on the $k_z$ composition of QW-states, is more fundamental compared to the $E(z)$-only effects (Jovic et al 2017, Moser et al 2018).

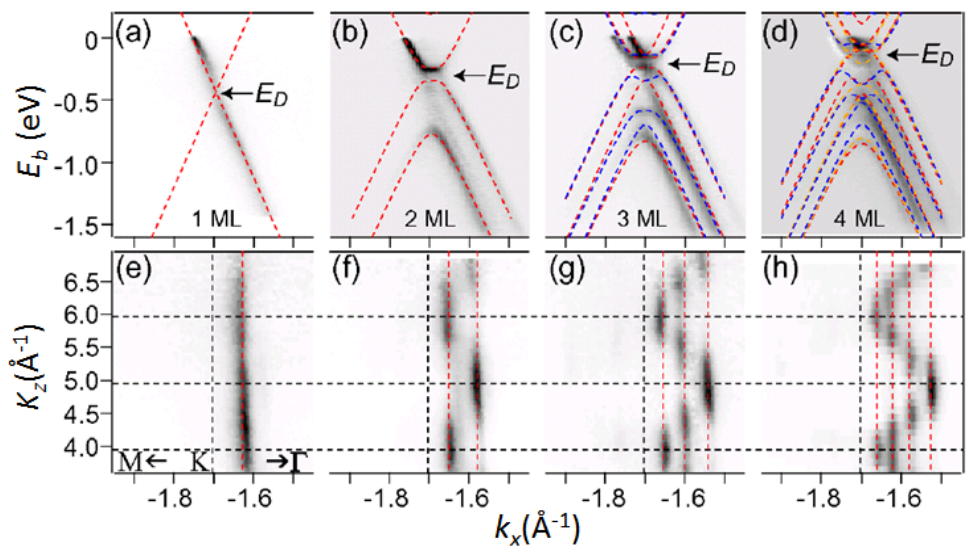

**Fig. 3**. ARPES data on graphene multilayers: (*a-d*) The π and π* 2D bands near $E_F$ for $m = 1 - 4$ graphene layers, respectively. The dashed lines are tight binding calculations; (*e-h*) Corresponding $I_{PE}$ from the π bands at $E_F - 1$eV as a function of $(k_x, K_z)$. With increase of $m$ these oscillations gradually approach a $K_z$-dispersive pattern characteristic of the 3D states (adapted from Ohta et al 2007).

An impressive example of ARPES applied to QW-states formed in buried heterostructures is the recent study by Lev et al (2018) on the GaAlN/GaN interface (for entries on nitride heterostructures see Medjdoub & Iniewski 2015). A large conduction band (CB) offset augmented by strong electric polarization forms at this interface a deep that confines mobile 2D electron gas on the intrinsic GaN side. It typically embeds two QW-states whose Airy-like $E(z)$-functions, resulting from solution of the model Poisson-Schrödinger equations in the approximately triangular QW, are sketched in Fig. 4(*a*). The spatial separation of these QW-states from the defect-rich GaAlN barrier layer allows the electrons to escape defect scattering and thereby dramatically increase their mobility. This idea is the fundamental operational principle of so-called high electron mobility transistors (HEMTs) finding their applications in a wide range of microwave devices such as cell phones and radars (Mimura 2002). Because of a relatively large thickness of the GaAlN layer in the operational HEMT heterostructures (in the reported case ~3 nm) the buried QW-states can only be accessed with soft-

X-ray ARPES at *hv* above ~1000 eV delivering sufficient photoelectron λ (Strocov *et al* 2014). Below we will analyse the ARPES response aspects of the study by Lev *et al* (2018) whereas for its further aspects, including planar anisotropy of electron effective mass propagating into non-linear transport properties of the GaAlN/GaN heterostructures, the reader is referred to the original publication by Lev *et al* (2018).

Fig. 4(*b*) reproduces the experimental Fermi surface (FS) map of the buried QW-states measured as a function of the in-plane ($K_x$, $K_y$) coordinates at *hv* = 1057 eV (one of the $I_{PE}(hv)$ peaks). The FS appears as tiny electron pockets located at the $\bar{\Gamma}$-points of the 2D interfacial BZ. As we discussed above, their intensity distribution over different BZs reflects essentially the 2D Fourier series of the QW wavefunctions. The fact of non-vanishing $I_{PE}$ in higher BZs evidences that these wavefunctions embed higher-**g** Fourier components coming from the $B(z)$-part of the QC-wavefunctions, which is beyond their simplistic $E(z)$-only description. A momentum-distribution curve (MDC) at Fermi energy $E_F$ measured across the $\bar{\Gamma}_{11}$-point is displayed in (*c*). Its two external peaks are related to the Fermi momentum $k_F^1$ of the first QW-state (QWS$_1$) and two merging internal ones to $k_F^2$ of the second QW-state (QWS$_2$) extending deeper into the GaN bulk, as sketched in (*b*).

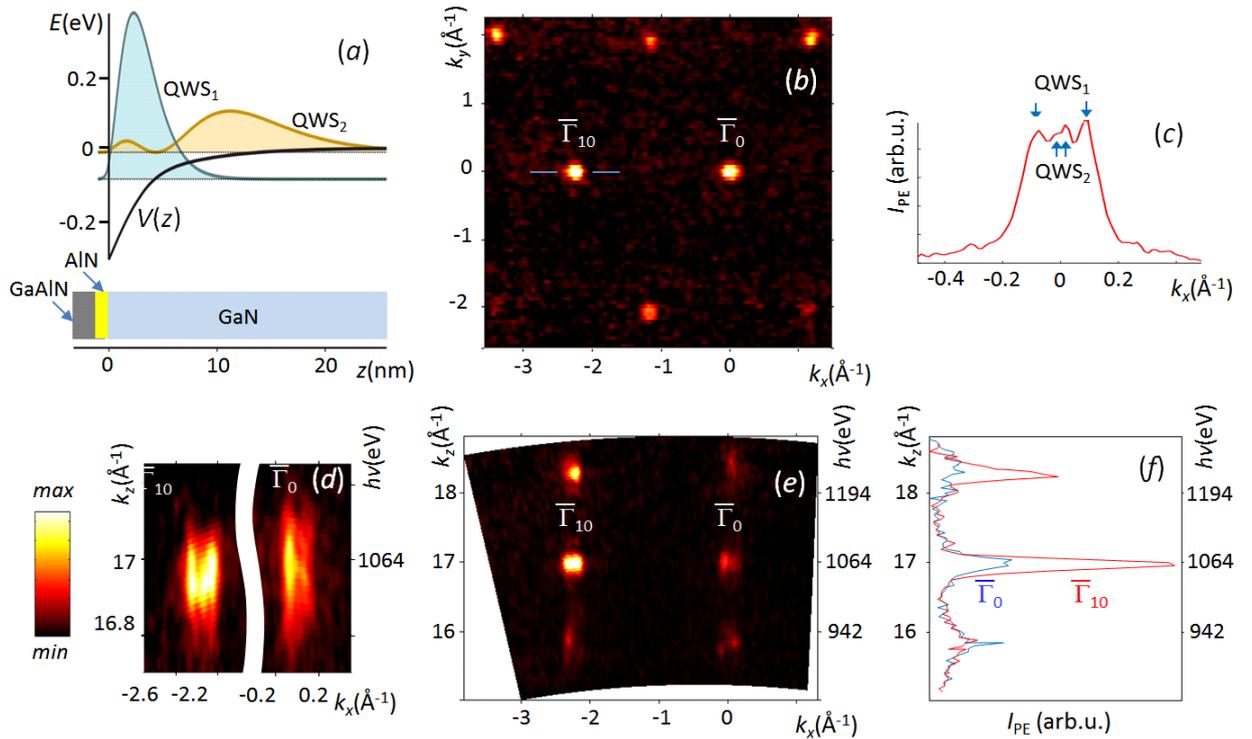

**Fig. 4.** ARPES data on the buried QW-states in AlGaN/GaN heterostructure: (*a*) Model $E(z)$-function of the two QW-states (top) embedded in the AlGaN/GaN interface (bottom); (*b*) Experimental in-plane FS measured at *hv* = 1057 eV; (*c*) MDC at $E_F$ across the $\bar{\Gamma}_{10}$-point (marked in *b*) whose peaks identify the two QW-states; (*e*) Experimental FS cross-section in the $\bar{\Gamma M}$ azimuth as a function of $K_z$ and (*d*) its zoom-in. The 2D character of the QW-states manifests itself in absence of their $K_z$-dispersion; (*f*) Fermi intensity in vicinity of the $\bar{\Gamma}_0$- and $\bar{\Gamma}_{10}$-points as a function of $K_z$. The ARPES response peaks coinciding with the integers of $G_z$ confirm that these QW-states are derived from the CBM of bulk GaN in the Γ-point (adapted from Lev *et al* 2018).

Most importantly, Fig. 3(*e*) reproduces the experimental FS of the QW-states measured in the ($K_x$,$K_z$) coordinates. The 2D character of these states manifests itself by the straight FS contours without any $K_z$-dispersion as emphasized by the zoom-in (*d*). As predicted by our FT-based theory, the ARPES response $I_{PE}(K_z)$ shows periodic peaks whenever $K_z$ hits the Γ-points (integers of $G_z$) in the out-of-plane direction. This is particularly clear in the line plots (*f*) showing Fermi intensity of the two QW-states integrated within ±$k_F{}^2$ around the $\bar{\Gamma}_0$- and $\bar{\Gamma}_{10}$-points ($k_z$-separation between the two ±$k_z$ Bloch-wave components in these QW-states is negligible). Although $K_z$-width of the $I_{PE}(K_z)$ peaks is limited in this *hv*-range mostly by the final-state Δ$k_z$, in the zoom-in (*d*) we can distinguish somewhat reduced $k_z$-width of the QWS$_2$ compared to QWS$_1$ which is consistent with the larger spatial extension of the former. Importantly, the $K_z$-foci of the ARPES peaks coincide with the Γ-points of the bulk BZ. This fact identifies the QW-states in the AlGaN/GaN heterostructures as originating from the CB minimum (CBM) of the three-dimensional band structure of bulk GaN. In a methodological perspective, such identification can be essential, for example, for studies of valleytronics materials (for entries see Schaibley *et al* 2016) where the CB can include several local minima almost degenerate in energy but separated in **k**.

*LO-states*

ARPES response of the LO-states, aperiodic and decaying with $K_z$, can be illustrated, for example, by recent results of Weiss *et al* (2015) on a monolayer of perylene tetracarboxylic dianhydride (PTCDA) absorbed on Ag(110). The PTCDA molecular orbitals form LO-type states whose $I_{PE}(K_z)$ is described within the above Fourier-transform formalism. Fig. 5 reproduces the Weiss' *et al* experimental $I_{PE}(hv)$ in the central photoemission lobe (points on the graph) for the HOMO and LUMO electron states formed by $p_z$-orbitals. These dependences are aperiodic and decay with increase of *hv*, in agreement with the above theoretical predictions. The zero value of $I_{PE}(hv)$ at $K_z$ = 0 reflects, as mentioned above, the out-of-plane antisymmetric character of the $p_z$-orbitals.

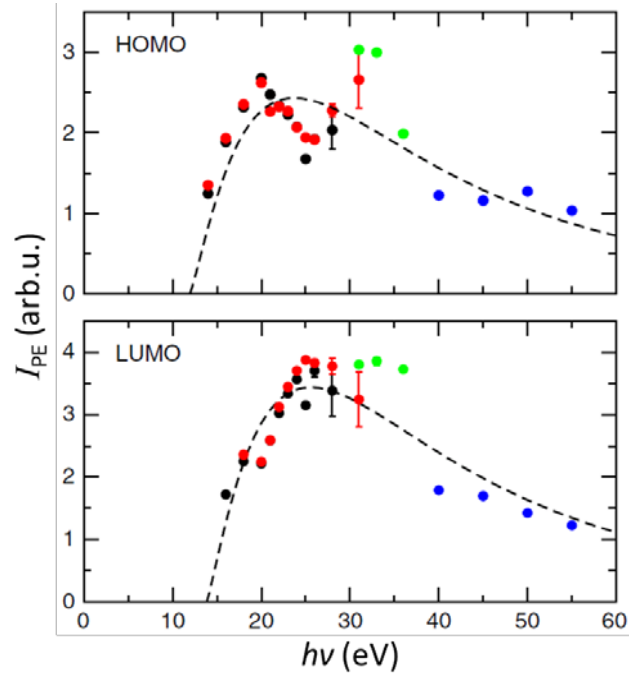

**Fig. 5**. *hv*-dependent ARPES intensity from the HOMO and LUMO states in a PTCDA molecular monolayer. The aperiodic dependence decaying with increase of *hv* is characteristic of the LO-states. Deviations from predictions of the FE-approximation (dashed lines) are attributed to the shape resonances (adapted from Weiss *et al* 2015).

Deviations of experimental points in Fig. 5 from the smooth trend (dashed lines) expected from FE-description of the final states are attributed to so-called shape resonances. This phenomenon originates from large variations of *V*(**r**) at the edges of the adsorbed molecules which distort the ideal FE-wavefront of the final states (Weiss *et al* 2015). We note that 3D-reconstruction of molecular orbitals with ARPES (orbital tomography) can be quite sensitive to such non-FE effects (for a critical review see Bradshaw & Woodruff 2015) because these experiments, restricted by the $I_{PE}$ decay with *hv* and by photo-induced damage of the molecules, typically operate at relatively low energies (below ~100 eV) where $E_k$ can be comparable with relatively strong *V*(**r**)-variations between the adsorbed molecules and at their interface with the substrate.

Returning to the conventional solid-state systems, the $I_{PE}$ decay with $K_z$ characteristic of the LO-states readily explains, for example, the empirical observation that increase of *hv* clears the ARPES spectra from stray intensity originating, for example, from surface defects left behind by cleavage of the sample (Razzoli *et al* 2012) or introduced by an amorphous overlayer (Woerle 2017).

## Conclusion

A lucid Fourier-transform based formalism has been presented that directly relates *hv*-dependent ARPES response of 2D states (surface, interface and QW-states) to Fourier composition and spatial confinement of their wavefunctions. For the QC-type states (including Shockley-Tamm type surface and interface states, and QW-states) formed by out-of-plane periodic Bloch waves modulated by envelope $E(z)$-functions, the *hv*-dependent ARPES response shows a characteristic pattern of periodic peaks located where photoelectron momentum matches harmonics of the QC-wavefunctions. Amplitudes of the peaks represent the corresponding Fourier series, and their broadening spatial extension of the $E(z)$-function. For the LO-type states of purely surface or interface origin (dangling-bond type and defects) the *hv*-dependent ARPES response is described by an aperiodic Voigt-type profile rapidly decaying with *hv*, related to spatial localization of these states in the out-of-plane direction. This transparent relation between wavefunctions and their ARPES response enables detailed analysis of different surface, interface and QW-states, as illustrated with ARPES data on the Al(100) surface state, buried interface state in GaAlN/GaN heterostructures, molecular orbitals of PTCDA, etc. This analysis is particularly accurate in the soft-X-ray energy range where the final states are close to pure plane waves with small intrinsic out-of-plane momentum broadening. The developed methodology paves ways towards experimental analysis and design of 2D wavefunctions in real electronic devices.

## Acknowledgements

I thank E. E. Krasovskii for sharing theoretical concepts and critical reading of the manuscript, J. H. Dil for insightful discussions, and L. L. Lev for promoting scientific exchange and help with the GaAlN/GaN data processing. The permissions of P. Hofmann, E. Rotenberg and P. Puschnig to reproduce their impressive results are gratefully acknowledged.